\documentclass[aps,pre,twocolumn,showpacs,floatfix]{revtex4} 
\usepackage{amsmath,amssymb,isolatin1,graphicx,times}

\newcommand{\be}{\begin{equation}}
\newcommand{\ee}{\end{equation}}
\newcommand{\bea}{\begin{eqnarray}}
\newcommand{\eea}{\end{eqnarray}}
\newcommand{\bw}{\begin{widetext}}
\newcommand{\ew}{\end{widetext}}

\newcommand{\kommentar}[1]{}

\begin{document}

\title{Inefficient quantum walks on networks: the role of the density
of states}
\author{Oliver M{\"u}lken}
\affiliation{
Theoretische Polymerphysik, Universit\"at Freiburg,
Hermann-Herder-Straße 3, 79104 Freiburg i.Br., Germany}

\date{\today} 
\begin{abstract}
We show by general arguments that networks whose density of states
contains few highly degenerate eigenvalues result in inefficient
performances of continuous-time quantum walks (CTQW) over these networks,
while systems whose eigenvalues all have the same degeneracy lead to very
efficient transport. We exemplify our results by considering CTQW and, for
comparison, its classical counterpart, continuous-time random walks, over
simple structures, whose eigenvalues and eigenstates can be calculated
analytically. Extensions to more complicated, hyper-branched networks are
discussed.
\end{abstract}
\maketitle

\section{Introduction}

The transfer of information in quantum systems has attracted a lot of
attention in recent years, especially in the context of quantum computing
\cite{kempe2003}.  In analogy to classical random walks, which are used as
algorithmic tool in ``classical'' computing, two version of quantum walks
have been introduced: discrete-time quantum (random) walks, with an
additional internal ``coin'' degree of freedom \cite{aharonov1993}, and
continuous-time quantum walks (CTQW), where the analogy to continuous-time
random walks (CTRW) lies in identifying the classical transfer matrix with
the quantum mechanical Hamiltonian \cite{farhi1998}.  Recently, it has
been shown how these two version are related \cite{strauch2006}. 

CTQW are formally equivalent to the tight-binding model in solid-state
physics \cite{Ziman} or the H{\"u}ckel/LCMO model in physical chemistry
\cite{McQuarrie} and therefore can be applied to study transport processes
in various types of different systems, like spin chains \cite{bose2003} or
ultra-cold Rydberg gases \cite{mbagrw2007}. What matters is that the
constituting elements (spins, atoms, molecules, etc.) are of the same
type, in the simplest cases they resemble two-level systems. 

In general, the quantum dynamics is very different from the corresponding
classical dynamics. The unitary time evolution of a quantum system leads
to characteristic dynamical phenomena, {\it e.g.}, quantum revivals and quantum
carpets for the particle in a box \cite{kinzel1995} or for CTQW on a ring
\cite{mb2005b}, or, for disordered systems, to localization
\cite{anderson1958}.  

For linear (ordered) systems, the quantum transport efficiency of CTQW has
been proven to overcome the classical efficiency, a result which
translates also to more complicated systems like decision or Cayley trees
\cite{childs2002}.  However, the efficiency of the transport strongly
depends on the initial condition \cite{mb2005a}, {\it i.e.}, different initial
conditions lead to vastly different dynamics of the CTQW.  

One way of quantifying the global efficiency of quantum walks is by the
{\sl average} probability of a walker to return to or stay at the origin
\cite{mb2006b}.  In the classical case this quantity depends only on the
eigenvalues, or more generally on the density of states (DOS), of the
underlying system and {\sl not} on its eigenvectors. Quantum mechanically,
there exists a lower bound to the quantum mechanical return probability.
This lower bound also depends only on the DOS \cite{mbb2006a,mb2006b},
and, in most cases considered here, is a good measure of the transport
efficiency, since the global temporal behaviour of the lower bound is
similar to the one of the full expression for the return probability which
also requires the eigenstates.

Depending (mainly) on the topology of the system, the DOS shows very
distinct features. There is a large variety of complex classical systems,
ranging from glasses to proteins, showing anomalous transport
\cite{metzler2000}, which can be related to the structure of the DOS. For
instance, CTRW over small-world networks have been
shown to be super-diffusive \cite{jespersen2000}.  As we proceed to show,
for quantum walks especially the degeneracies of the eigenvalues and the
number of degenerate eigenvalues determine the transport efficiency.

\section{Quantum walks on networks}\label{sec_qw}

We start by considering quantum mechanical transport processes on discrete
networks, which are a collection of $N$ connected nodes.  A connectivity
matrix ${\bf A} = (A_{ij})$ can be assigned to every network. The
non-diagonal elements $A_{ij}$ equal $-1$ if nodes $i$ and $j$ are
connected by a bond and $0$ otherwise. The diagonal elements $A_{ii}$
equal the number of bonds, $f_i$, which exit from node $i$.

\subsection{Transition probabilities}

Classically, a CTRW is governed by a master equation for the conditional
probability, $p_{k,j}(t)$, to find the walker at time $t$ at node $k$ when
starting at node $j$.\cite{weiss} The transfer matrix of the walk, ${\bf
T} = (T_{kj})$, is, in the simplest case where the transmission rates
$\gamma$ of all bonds are take to be equal, related to the connectivity
matrix by ${\bf T} = - \gamma {\bf A}$ (we assume $\gamma\equiv 1$ in the
following).

CTQWs are obtained by identifying the Hamiltonian of the system with the
classical transfer matrix, ${\bf H} = - {\bf T}$
\cite{farhi1998,childs2002,mb2005a}.  The states $|j\rangle$ associated
with the nodes $j$ of the network form a complete, ortho-normalised basis
set of the whole accessible Hilbert space, {\it i.e.}, $\langle k | j \rangle =
\delta_{kj}$. A state $| j \rangle$ evolves in time as $| j(t) \rangle =
{\bf U}(t) |j \rangle$, where ${\bf U}(t) = \exp(-{\rm i} {\bf H} t)$ is the
quantum mechanical time evolution operator (we have set $m\equiv1$ and
$\hbar\equiv1$). 

The transition amplitude $\alpha_{k,j}(t)$ from state $| j \rangle$ at
time $0$ to state $|k\rangle$ at time $t$ reads then $\alpha_{k,j}(t) =
\langle k | {\bf U}(t) | j \rangle$ and obeys Schr\"odinger's equation.
Denoting the eigenvalues of the Hamiltonian ${\bf H} = -{\bf T}$ by $E_n$
($n=1,\dots,N$) and the ortho-normalised eigenstates by $| \psi_n\rangle$,
such that $\sum_n | \psi_n\rangle \langle  \psi_n | = \boldsymbol 1$, the
quantum mechanical transition probability is
\begin{equation}
\pi_{k,j}(t) \equiv |\alpha_{k,j}(t)|^2 = \left| \sum_n \langle k|
{\rm e}^{-{\rm i} E_n t} | \psi_n\rangle \langle  \psi_n | j \rangle \right|^2.
\label{qm_prob_full}
\end{equation}

\subsection{Long time limit}

Quantum mechanically the unitary time evolution prevents $\pi_{k,j}(t)$
from having a definite limit for $t\to\infty$. In order to compare the
classical long time probability with the quantum mechanical one, one
usually uses the long time average (LTA) \cite{aharonov2001}
\begin{eqnarray}
\chi_{k,j} &\equiv& \lim_{T\to\infty} \frac{1}{T} \int_0^T {\rm d}t \
\pi_{k,j}(t), \label{limprob} \\
&=&
\sum_{n,m} \delta_{E_n,E_m} \langle k | \psi_n \rangle \langle \psi_n | j
\rangle \langle j | \psi_m \rangle \langle \psi_m | k \rangle,
\label{limprob_ev}
\end{eqnarray}
where $\delta_{E_n,E_m} = 1$ if $E_n=E_m$ and $\delta_{E_n,E_m} = 0$
otherwise.  Some eigenvalues of $\bf H$ might be degenerate, so that the
sum in eq.~(\ref{limprob_ev}) can contain terms belonging to different
eigenstates $| \psi_n\rangle$ and $| \psi_m\rangle$.

We can use the Cauchy-Schwarz inequality to obtain a lower bound for the
LTA \cite{mbb2006a}, such that the time integral in eq.~(\ref{limprob})
fulfills 
\begin{equation}
\int_0^T {\rm d}t \ |\alpha_{k,j}(t)|^2 \ge \frac{1}{T}\left| \int_0^T
{\rm d}t \
\alpha_{k,j}(t) \right|^2 .
\label{csi_alpha}
\end{equation}
This results in
\begin{equation}
\chi_{k,j} \ge \lim_{T\to\infty} \left| \frac{1}{T} \int_0^T {\rm d}t \ \sum_n
\langle k| {\rm e}^{-{\rm i} E_n t} | \psi_n\rangle \langle  \psi_n | j \rangle
\right|^2.
\label{csi_alpha1}
\end{equation}
The only term in the sum over $n$ in eq.~(\ref{csi_alpha1}) which survives
after integration and taking the limit $T\to\infty$ is the one with
$E_1=0$.  The corresponding eigenvector can be written as $|\psi_1\rangle
= 1/\sqrt{N}\sum_{j=1}^{N} | j\rangle$ \cite{mbb2006a}. Since $\langle
k|\psi_1\rangle = \langle \psi_1 | j\rangle = 1/\sqrt{N}$ we get with
eq.~(\ref{csi_alpha1})
\begin{equation}
\chi_{k,j} \ge \left| \langle k | \psi_1 \rangle \langle \psi_1 | j
\rangle \right|^2 = \frac{1}{N^2}.
\label{lbchi}
\end{equation}

\subsection{Averaged transition probabilities}

Quantum mechanically as well as classically, we can calculated properties
from the probability distributions, which solely depend on the eigenvalues
of ${\bf H}$ or ${\bf T}$, respectively, but {\sl not} on the eigenstates.
Classically, there exists a simple expression for the {\sl average} return
probability to the initially excited node (see, {\it e.g.},
ref.~\cite{blumen2005}):
\begin{equation}
\overline{p}(t) \equiv \frac{1}{N} \sum_{j=1}^{N} p_{j,j}(t) = \frac{1}{N}
\sum_{n=1}^{N} \ {\rm e}^{-E_n t}.
\label{pclavg}
\end{equation}
Quantum mechanically, the average is given by
\begin{equation}
\overline{\pi}(t) \equiv \frac{1}{N} \sum_{j=1}^{N} \pi_{j,j}(t).
\label{pqmavg0}
\end{equation}
By using the Cauchy-Schwarz inequality we obtain a lower bound for
$\overline{\pi}(t)$ \cite{mbb2006a,mb2006b},
\begin{equation}
\overline{\pi}(t) = \frac{1}{N^2} \sum_{j=1}^{N} |\alpha_{j,j}|^2
\sum_{l=1}^{N} 1 \geq \left| \frac{1}{N} \sum_j \alpha_{j,j} \right|^2
\equiv |\overline{\alpha}(t)|^2.
\end{equation}
With eq.~(\ref{qm_prob_full}) we therefore get
\begin{equation}
\overline{\pi}(t) \geq \left|\frac{1}{N} \sum_{n} \ {\rm e}^{-{\rm i}E_nt}\right|^2 .
\label{pqmavg}
\end{equation}
In analogy to the classical case, the lower bound
$|\overline{\alpha}(t)|^2$ depends only on the eigenvalues and {\sl not}
on the eigenstates of ${\bf H}$. 

\section{Quantum walk efficiencies}\label{sec_eff}

Equations~(\ref{pclavg}) and (\ref{pqmavg}) can be used to quantify the
efficiency of the two transport processes \cite{mb2006b}: In the classical
case a quick decrease of $\overline{p}(t)$ must result - on average - in a
quick increase of the probability for the walker to be at any other but
the initial node.  Thus, the transport away from the initial node $j$ is
the more efficient the quicker the decrease of $\overline{p}(t)$. For
instance, it has been shown for small-world networks that
$\overline{p}(t)$ follows a stretched exponential rather than a power law
as for regular networks, which gives rise to a quicker decrease of
$\overline{p}(t)$ and thus to a super-diffusive behaviour
\cite{jespersen2000}. Quantum mechanically, $\overline{\pi}(t)$, as well
as the lower bound $|\overline{\alpha}(t)|^2$, will show strong
oscillations due to the unitary time evolution. However, one can use the
envelope of this oscillations as a measure for the efficiency, for which
the same arguments as in the classical case apply. For regular
$d$-dimensional networks it is straightforward to show that the envelope
of $|\overline{\alpha}(t)|^2$ decays as $t^{-d}$, whereas
$\overline{p}(t)$ decays as $t^{-d/2}$ \cite{mb2006b}. Since the decay of
$|\overline{\alpha}(t)|^2$ is much quicker than the one of
$\overline{p}(t)$, the quantum walk is more efficient than the classical
random walk.

For finite systems, $\overline{p}(t)$, $\overline{\pi}(t)$, and
$|\overline{\alpha}(t)|^2$ do not decay ad infinitum.  Classically, the
averaged probability will drop in the course of time to the equipartition
value of $1/N$, since we always have the eigenvalues $E_1=0$ which is
non-degenerate, {\it i.e.},
\begin{equation}
\overline{p}(t) = \frac{1}{N} + \frac{1}{N} \sum_{n=2}^N \ {\rm e}^{-E_n t}.
\end{equation}
Quantum mechanically, both, $\overline{\pi}(t)$ and
$|\overline{\alpha}(t)|^2$, will be oscillating about a value given by the
LTA of $\overline{\pi}(t)$ or its lower bound $|\overline{\alpha}(t)|^2$,
respectively.  The long time average of $\overline{\pi}(t)$ follows as
\begin{equation}
\overline{\chi} \equiv \lim_{T\to\infty}\frac{1}{T} \int_0^T {\rm d}t \
\overline{\pi}(t) = \frac{1}{N}  \sum_{j,n} \big|\langle j | \psi_{n}
\rangle \big|^4,
\label{chi_exact}
\end{equation}
which still depends on the eigenstates $|\psi_n\rangle$. Using again
the Cauchy-Schwarz inequality to obtain a lower bound for
$\overline{\chi}$ we get with eq.~(\ref{pqmavg})
\begin{equation}
\overline{\chi} \geq \frac{1}{N^2} \sum_{n,m} \delta_{E_n,E_m}.
\label{chi_lowerbound}
\end{equation}
These long time averages give indications on the overall performance of
quantum walks: If the LTA are larger than the equipartition value of
$1/N$, on average most of the probability will remain at the initial node.

The argument of using the lower bound $|\overline{\alpha}(t)|^2$ instead
of $\overline{\pi}(t)$ goes only one way: Only if the envelope of
$|\overline{\alpha}(t)|^2$ exceeds the one of the classical probability
$\overline{p}(t)$ at any(most) times is the quantum walk less efficient
than the classical random walk. However, if the envelope of
$|\overline{\alpha}(t)|^2$ lies below $\overline{p}(t)$, in general,
little can be said about $\overline{\pi}(t)$. It might also happen that
the envelope $\overline{\pi}(t)$ lies above the classical curve, although
in the examples considered below this is not the case.

\section{Degeneracies of eigenvalues}\label{sec_dos}

The eigenvalues of a large variety of networks will be degenerate. By
denoting the degeneracy of $E_n$ by $D_n\equiv D(E_n)$, we can recast
eqs.~(\ref{pclavg}) and (\ref{pqmavg}) into
\begin{eqnarray}
\overline{p}(t) &=& \frac{1}{N} \sum_{E_n} \ D_n \ {\rm e}^{-E_n t}\\
\overline{\pi}(t) &\geq& \left|\frac{1}{N} \sum_{E_n} \ D_n \
{\rm e}^{-{\rm i}E_nt}\right|^2 = |\overline{\alpha}(t)|^2,
\end{eqnarray}
respectively. As we will show, there is a profound difference between the
temporal behaviour of the two quantities depending on the eigenvalues and
their degeneracy. For classical transport processes, the long time
behaviour is dominated by the smallest eigenvalues, no matter what their
degeneracy is - large eigenvalues with high degeneracies will eventually
be suppressed by the exponential. In the quantum case, however, the
degeneracies become important due to the unitary time evolution.

\subsection{Uniform degeneracy of eigenvalues}\label{sec_uniform}

We start with networks, whose eigenvalues have degeneracies of the order
${\cal O}(1)$.  Then we rewrite the lower bound as
\begin{equation}
\left|\overline{\alpha}(t)\right|^2 = \frac{2}{N^2} \sum_{E_n,E_m>E_n} \
D_n D_m \ \cos[(E_n-E_m)t].
\label{alpha_nondeg}
\end{equation}
Now, the product $D_n D_m$ will be of order ${\cal O}(1)$, therefore, all
terms in the sums contribute to $|\overline{\alpha}(t)|^2$. The shape of
the decay depends on the functional form of the degeneracy. If we assume
$D_n \sim E_n^{-\nu}$, we obtain $\overline{p}(t) \sim t^{-\nu}$,
whereas $\left|\overline{\alpha}(t)\right|^2 \sim t^{-2\nu}$
\cite{mb2006b}.   
At large times
$t$ the only terms which contribute to the sum are those for which
$E_n=E_m$. Then, the sum will be of order ${\cal O}(N)$ and therefore
$\left|\overline{\alpha}(t)\right|^2$ will be of order ${\cal O}(1/N)$. 

\begin{figure}[htb]
\centerline{\includegraphics[clip=,width=\columnwidth]{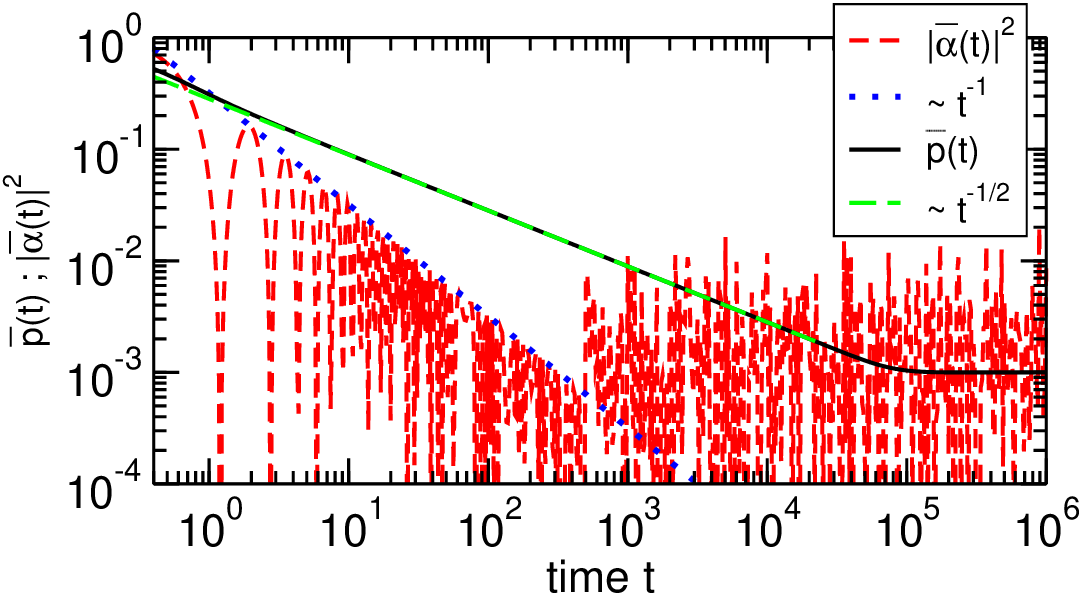}}
\caption{(Colour on-line) $\overline{p}(t)$ and
$|\overline{\alpha}(t)|^2$ with the appropriate scaling
$t^{-1/2}$ and $t^{-1}$, respectively, for a ring of size $N=1000$. }
\label{ring} 
\end{figure}

As an example, fig.~\ref{ring} shows $\overline{p}(t)$ (solid black line)
and $|\overline{\alpha}(t)|^2$ (short dashed red line) for a regular network of
$N=1000$ nodes. For regular networks, almost all eigenvalues are two-fold
degenerate, for odd(even) $N$ there is one(two) non-degenerate eigenvalue.
Note that the lower bound is exact in this case, {\it i.e.},
$|\overline{\alpha}(t)|^2 = \overline{\pi}(t)$ \cite{bbm2006a}.

While propagating without interference, the envelope of
$\left|\overline{\alpha}(t)\right|^2$ decays as $t^{-1}$ (dotted blue
line), which is faster than in the classical case, where $\overline{p}(t)
\sim t^{-1/2}$ (long dashed green line) \cite{mb2006b}. Therefore, the
quantum walk can be considered to be more efficient than the corresponding
classical random walk.  After the interference sets in at about
$t\approx500=N/2$, $\left|\overline{\alpha}(t)\right|^2$ fluctuates about
the LTA $\overline{\chi} = (2N-1)/N^2$ [$\overline{\chi} = (2N-2)/N^2$]
for odd(even) $N$, which is indeed of the same order as the classical long
time equipartition value $1/N$. However, the classical plateau value is
reached at a much later time at about $t\approx 10^5$. 

\subsection{One highly degenerate eigenvalue}

In contrast to the uniform degeneracy, we consider now one eigenvalue,
$E_l$, whose degeneracy, $D_l$, is of order ${\cal O}(N)$ whereas the
others are of order ${\cal O}(1)$ or less. By writing
\begin{equation}
\overline{\alpha}(t) = \frac{1}{N}\Bigg[ D_l \  {\rm e}^{-{\rm i}E_lt} + \sum_{E_n\neq
E_l} \ D_n \ {\rm e}^{-{\rm i}E_nt} \Bigg]
\end{equation}
we obtain, up to order ${\cal O}(1/N^2)$,
\begin{equation}
\left| \overline{\alpha}(t) \right|^2 \approx \frac{D_l}{N^2} \Bigg\{ D_l
+ \sum_{E_n\neq E_l} D_n \ 2\cos[(E_l - E_n) t] \Bigg\}.
\label{alpha_deg}
\end{equation}
The first term on the right-hand side of eq.~(\ref{alpha_deg}) is of order ${\cal
O}(1)$ and the second term [with the same argument as for
eq.~(\ref{alpha_nondeg})] is of order ${\cal O}(1/N)$. Therefore, for few
highly degenerate eigenvalues, the lower bound $\left|
\overline{\alpha}(t) \right|^2$ will not show a decay to values which
fluctuate about $1/N$ but rather will fluctuate for all times about
$1-1/N$. Also $\overline{\pi}(t)$ will not decay but fluctuate roughly
about the same value since $\left| \overline{\alpha}(t) \right|^2$ is a
lower bound.

\begin{figure}[htb]
\centerline{\includegraphics[clip=,width=\columnwidth]{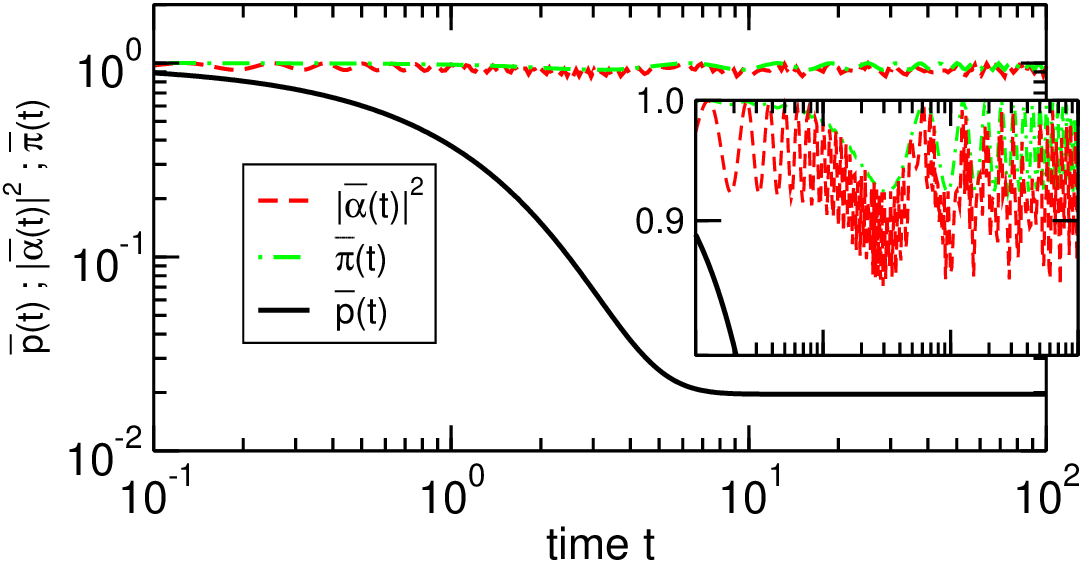}}
\caption{(Colour on-line). $\overline{p}(t)$, $\overline{\pi}(t)$, and
$|\overline{\alpha}(t)|^2$ for a star with $N=51$ nodes. The inset shows a
close-up of $\overline{\pi}(t)$ and $|\overline{\alpha}(t)|^2$ in the same
time interval.
}
\label{star1} 
\end{figure}

Take as a simple example a star-shaped network, having one core node and
$N-1$ nodes directly connected to the core but not to each other. The
eigenvalue spectrum has a very simple structure, there are $3$ distinct
eigenvalues, namely $E_1=0$, $E_2=1$, and $E_3=N$, having the degeneracies
$D_1=1$, $D_2=N-2$, and $D_3=1$, respectively.  Therefore we get
\begin{eqnarray}
\overline{p}(t) &=& \frac{1}{N} \left[ 1 + (N-2) {\rm e}^{-t} + {\rm e}^{-(N-2)t}
\right] \\
|\overline{\alpha}(t)|^2 &=& \frac{1}{N^2} \left| 1 + (N-2) {\rm e}^{-{\rm i}t} +
{\rm e}^{-{\rm i}(N-2)t} \right|^2.
\label{pi_star}
\end{eqnarray}
Obviously, only the term $|(N-2)\exp(-{\rm i}t)|^2/N^2 = (N-2)^2/N^2$ in eq.\
(\ref{pi_star}) is of order ${\cal O}(1)$.  All the other terms are of
order ${\cal O}(1/N)$ or ${\cal O}(1/N^2)$ and, therefore, cause only
small oscillations (fluctuating terms) about or negligible shifts
(constant terms) from $(N-2)^2/N^2 \approx 1-1/N$.

Figure~\ref{star1} shows $\overline{p}(t)$, $\overline{\pi}(t)$, and
$|\overline{\alpha}(t)|^2$ for a star-shaped network with $N=51$ nodes.
Here, not only $|\overline{\alpha}(t)|^2$ (short dashed red line) but also
$\overline{\pi}(t)$ (long dashed green line) fluctuate about a value close
to one, see also the inset of fig.~\ref{star1}.  Classically,
$\overline{p}(t)$ (solid black line) decays to the equipartition value
$1/N$, although the exact form of the decay is different than for regular
networks.  Thus, the quantum walk
has - on average - a large probability to return (or stay) at the initial
site, and in this sense is less efficient than its classical counterpart. 

\subsection{Two highly degenerate eigenvalues}

The next step is having two highly degenerate eigenvalues, $E_l$ and
$E_m$, with degeneracies $D_l$ and $D_m$ of order ${\cal O}(N/2)$. This
changes the properties of $\left| \overline{\alpha}(t) \right|^2$. From
\begin{eqnarray}
\overline{\alpha}(t) &=& \frac{1}{N} \Bigg[ D_l \ {\rm e}^{-{\rm i}E_lt} + D_m
{\rm e}^{-{\rm i}E_mt} \nonumber \\
&& + \sum_{E_n\neq \{E_l,E_m\}} \ D_n \ {\rm e}^{-{\rm i}E_nt}\Bigg],
\end{eqnarray}
we get, up to order ${\cal O}(1/N)$,
\begin{equation}
\left| \overline{\alpha}(t) \right|^2 \approx \frac{1}{N^2} \Big\{ D_l^2 +
D_m ^2 + 2 D_l D_m \cos[(E_l-E_m)t] \Big\}.
\label{alpha_deg2}
\end{equation}
Here, the terms $D_l^2$, $D_m ^2$, and $D_l D_m$ are of order ${\cal
O}(N^2/4)$; the cosine fluctuates in the interval $[0,1]$. Therefore, in
contrast to eq.~(\ref{alpha_deg}) whose fluctuations are of the order
${\cal O}(1/N)$, eq.~(\ref{alpha_deg2}) will also fluctuate in the full
interval $[0,1]$.  As before, there is no envelope of $\left|
\overline{\alpha}(t) \right|^2$ which decays with time. 

\begin{figure}[htb]
\centerline{\includegraphics[clip=,width=\columnwidth]{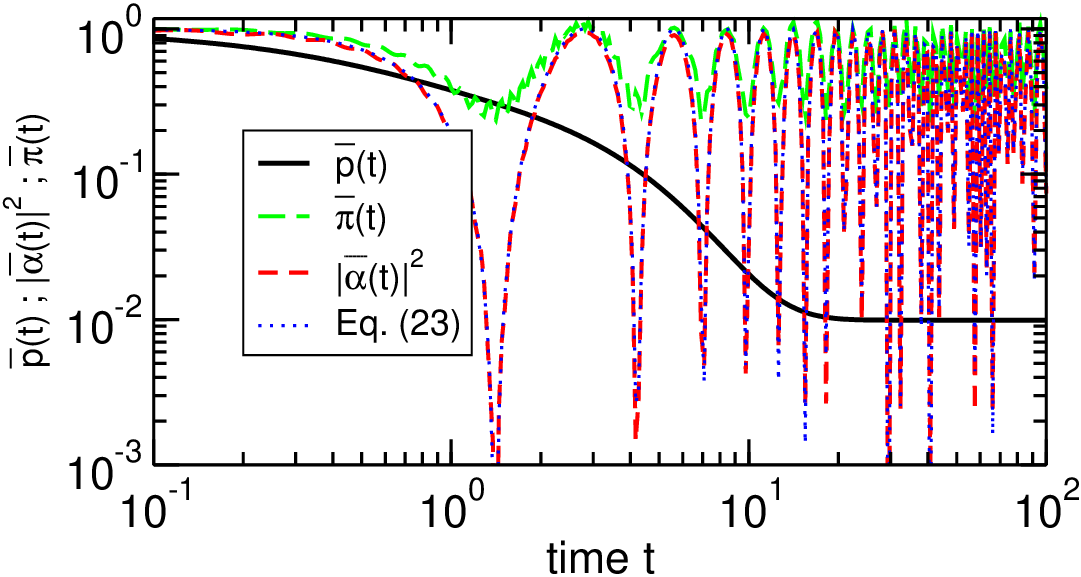}}
\caption{(Colour on-line) $\overline{p}(t)$, $\overline{\pi}(t)$, and
$|\overline{\alpha}(t)|^2$ for a star with $50$ arms of length $2$, {\it i.e.},
$N=101$ nodes.  
}
\label{star2} 
\end{figure}

An example of such a system is a star with $2$ nodes on each arm. There is
one zero eigenvalue $E_1=0$, two non-degenerate eigenvalues
$E_{2,3}=(N+5\pm\sqrt{N^2-6N+25})/4$, and two eigenvalues
$E_{4,5}=(3\pm\sqrt{5})/2$, whose degeneracy $D_4=D_5=(N-3)/2$, {\it i.e.}, it
is of order ${\cal O}(M)$, where $M=(N-1)/2$ is the number of arms. Then,
for large $N$, where we assume $N-3 \approx N$, eq.~(\ref{alpha_deg2})
becomes 
\begin{equation}
\left| \overline{\alpha}(t) \right|^2 \approx \frac{1}{2} \left[ 1 +
\cos(\sqrt{5} \ t)\right].
\label{alpha_star}
\end{equation}

Figure~\ref{star2} shows the temporal behaviour of $\overline{p}(t)$,
$\overline{\pi}(t)$, $|\overline{\alpha}(t)|^2$, and of
eq.~(\ref{alpha_star}) for a star with $50$ arms of length $2$, {\it
i.e.}, $N=101$ nodes. While the behaviour of $\overline{p}(t)$ (black
solid line) is similar to the one for the simple star (fig.~\ref{star1}),
there are obvious differences for $\overline{\pi}(t)$ (long dashed green
line) and $|\overline{\alpha}(t)|^2$ (short dashed red line). Although, as
for the star-shaped network, neither $\overline{\pi}(t)$ nor
$|\overline{\alpha}(t)|^2$ show a decay to some value comparable to the
classical equipartition value $1/N$, the oscillations become larger for
both quantities. While $\overline{\pi}(t)$ fluctuates roughly in the
interval $[0.2,1]$, $|\overline{\alpha}(t)|^2$ fluctuates in the full
interval $[0,1]$.  Moreover, considering only the two highly degenerate
eigenvalues results in an excelent agreement of eq.~(\ref{alpha_star})
(dotted blue line) with the full expression for
$|\overline{\alpha}(t)|^2$.  Thus also here the quantum walks have a large
probability to return or stay at the origin, which results in a low
transport efficiency compare to the classical random walk.

\subsection{Complex structures with highly degenerate
eigenvalues}\label{sec_complex}

So far we have considered simple systems. As we now turn to more complex
structure, it will become evident that a lot can be related to those
simple models. In particular, there are also large complex systems of
eminent interest which have only few degenerate eigenvalues. 

One group of systems of interest are star-like hyper-branched structures.
An example is the dendrimer, which has been experimentally realised as one
macromolecule and which has interesting applications, such as drug
delivery and as a light-harvesting antenna \cite{mukamel1997}. The nodes
of the dendrimer can be grouped into generations $g$, which are concentric
about the core node. In our case the core ($g=0$) has $3$ emanating bonds
connecting three additional nodes to it ($g=1$); every node in generation
$g\ge1$ is connected by one bond to two other node in generation $g+1$.
Therefore, total number of nodes grows exponentially with the total number
of generations $G$, {\it i.e.}, $N = 3\cdot 2^G -2$, whereas the number in
one generation $g$ is given by $N_g=3\cdot 2^{g-1}$. 

The eigenvalues of dendrimers can be calculated recursively
\cite{cai1997}, which yields one eigenvalue $E_1=0$ and $G+1$
non-degenerate eigenvalues.  Additionally, there are $N-G-2$ degenerate
eigenvalues, whose degeneracies increase with the number of generations.
In the first generation $g$, where the eigenvalue appears, it is two-fold
degenerate, while for larger generation the degeneracy grows as $3\cdot
2^{G-g-1}$. The first appearing degenerate eigenvalues are $E_2=1$ and
$E_{3,4}=2\pm\sqrt{3}$, whose degeneracies are $D_2=D_3=D_4 = 3 \cdot
2^{G-2}$ for $G>2$ and $3 \cdot 2^{G-3}$ for $G>3$.  Therefore, roughly
$3\cdot 3 \cdot 2^{G-3} \approx N/3$ eigenvalues, {\it i.e.}, $1/3$ of the total
number of eigenvalues, are given by $E_2$ and $E_{3,4}$. 

\begin{figure}[htb]
\centerline{\includegraphics[clip=,width=\columnwidth]{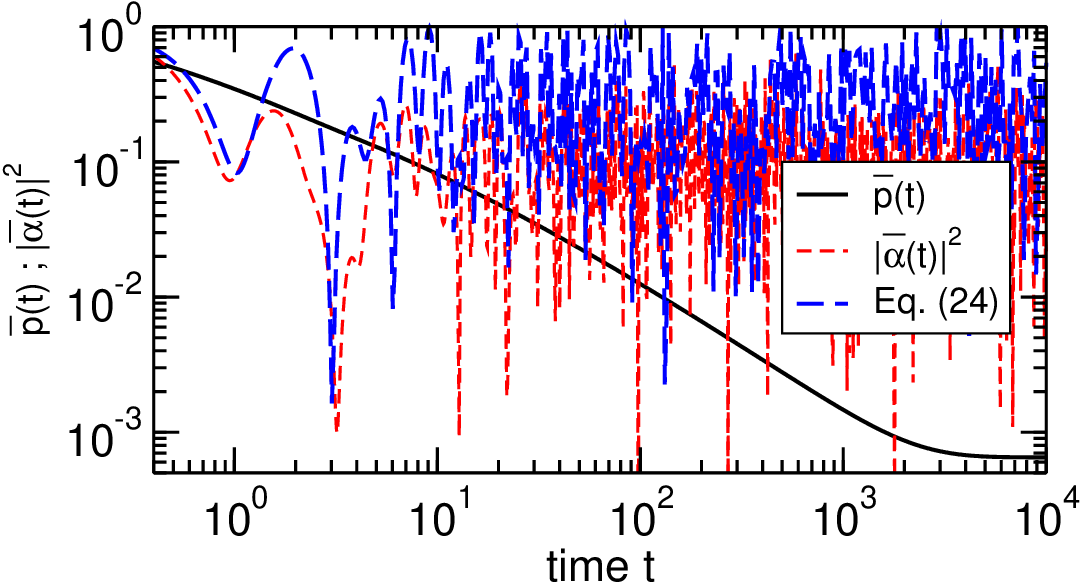}}
\caption{(Colour on-line) $\overline{p}(t)$, $|\overline{\alpha}(t)|^2$,
and the approximate value for $|\overline{\alpha}(t)|^2$
[eq.~(\ref{alpha_cayley})] for a dendrimer of generation $G=10$, {\it i.e.},
with $N=3070$ nodes.
} 
\label{dendrimer} 
\end{figure}

Figure~\ref{dendrimer} shows $\overline{p}(t)$ (solid black line) and
$|\overline{\alpha}(t)|^2$ (short-dashed red line) for a dendrimer of
generation $G=10$, {\it i.e.}, with $N=3070$ nodes.  Classically,
$\overline{p}(t)$ decays in a similar fashion as in figs.~\ref{star1} and
\ref{star2}. In the quantum case, the lower bound
$|\overline{\alpha}(t)|^2$ strongly oscillates about a value which is
close to $|\overline{\alpha}(t)|^2\approx0.2$. We stress again that
$|\overline{\alpha}(t)|^2$ is the lower bound to $\overline{\pi}(t)$ and
especially reproduces the maxima of $\overline{\pi}(t)$ quite well
\cite{mbb2006a}. Thus, also
$\overline{\pi}(t)\geq|\overline{\alpha}(t)|^2$ will be restricted to even
higher values, see also fig.~8 in ref.~\cite{mbb2006a}.

When restricting ourselves to only the three most highly degenerate
eigenvalues, the dendrimer resembles a star-like structures. From the
eigenvalues we obtain (for $G>3$)
\begin{equation}
|\overline{\alpha}(t)|^2 \approx \frac{1}{\cal N} \Bigg\{1 +
4\cos(\sqrt{3}t)\left[\cos(\sqrt{3}t) + \cos(t)\right] \Bigg\},
\label{alpha_cayley}
\end{equation}
where ${\cal N}$ is an appropriate normalization constant.
Figure~\ref{dendrimer} also shows the temporal behaviour of
eq.~(\ref{alpha_cayley}), see long-dashed blue line, which is very similar
to the full expression for $|\overline{\alpha}(t)|^2$.  Hence, the general
behaviour is indeed dominated by the most degenerate eigenvalues, there
are only slight deviations due to the remaining ones.  From
$|\overline{\alpha}(t)|^2$ as well as from eq.~(\ref{alpha_cayley}), we
find also here that quantum walks over dendrimers are much less efficient
than their classical counterpart.

There is certainly a large variety of networks, on which the quantum
mechanical transport behaviour remains to be investigated. Prime examples
are fractals. Classically it has been shown that the return probability
depends on the spectral (fracton) dimension of the fractal, see, {\it e.g.},
\cite{alexander1981}. The eigenvalues of certain fractals can also be
calculated recursively \cite{blumen2003}. In some cases this gives rise to
highly degenerate eigenvalues.


\kommentar{
Nevertheless, we saw in all examples considered above that especially
the maxima of $\overline{\pi}(t)$ are well reproduced by
$|\overline{\alpha}(t)|^2$ and, thus, the envelope of $\overline{\pi}(t)$
will follow the envelope of $|\overline{\alpha}(t)|^2$.

The eigenstates can change this behaviour. This change becomes most evident
in the LTA of $\overline{\pi}(t)$, eq.~(\ref{chi_exact}), and of its lower
bound $|\overline{\alpha}(t)|^2$, eq.~(\ref{chi_lowerbound}).

It has been shown, that for small-world networks, which were composed from
a ring by randomly adding new bonds between the nodes, the ensemble average
of the LTA
$\overline{\chi}$ increases with increasing the number of additional
bonds \cite{mpb2007a}. 
Although there is no localization for small-world
networks, the effect of increasing $\overline{\chi}$ should be
visible, for instance, for the Anderson model. There, the eigenstates
are highly localised which gives rise to large contribution to the sums in
eq.~(\ref{chi_exact}). Then the argument of using the lower bound for
measuring transport efficiency breaks down, but then there are also other
means to quantify transport processes \cite{anderson1958}.
}

\section{Conclusions}

We have analyzed the efficiencies of quantum walks over different discrete
networks by relating the density of states to the average return
probability of the walk. Depending on the connectivity of the networks,
the spectra of the Hamiltonians are vastly differerent. The difference between
classical random and quantum walks over networks can be summarised as
follows: In the classical case the eigenvalues of the transfer matrix
itself are governing the average return probabilities $\overline{p}(t)$.
Quantum mechanically, the degeneracies of the eigenvalues of the
Hamiltonian dominate the temporal behaviour and the long time averages of
the average return probability $\overline{\pi}(t)$ and its lower bound
$|\overline{\alpha}(t)|^2$.  After giving general arguments corroborating
this statements, we illustrated this by simple examples, like regular and
star-like networks and extended those to  more complex structures, like
Cayley trees or dendrimers, which also have few highly degenerate
eigenvalues. 

\acknowledgements
I thank Alexander Blumen for many illuminating discussion and his
continuous encouragement. Support from the Deutsche
Forschungs\-ge\-mein\-schaft (DFG), the Fonds der
Chemischen Industrie and the Ministry of Science, Research and the Arts of
Baden-W\"urttemberg (AZ: 24-7532.23-11-11/1) is gratefully acknowledged.

\end{document}